\documentclass[aps,prb,twocolumn,showpacs,amsmath,amssymb,superscriptaddress]{revtex4}
\usepackage{dcolumn}
\usepackage{bm}
\usepackage{graphicx}
\usepackage{times}
\usepackage{epstopdf}
\begin{document}

\renewcommand{\section}[1]{{\par\it #1.---}\ignorespaces}

\title{Strong potential impurities on the surface of a topological insulator}
\author{Annica M. Black-Schaffer}
 \affiliation{Nordic Institute for Theoretical Physics (NORDITA), Roslagstullsbacken 23, S-106 91 Stockholm, Sweden}
 \affiliation{Department of Physics and Astronomy, Uppsala University, Box 516, S-751 20 Uppsala, Sweden}
 \author{Alexander V. Balatsky}
\affiliation{Theoretical Division and Center for Integrated Nanotechnologies, Los Alamos National Laboratory, Los Alamos, New Mexico 87545, USA}

\begin{abstract}
Topological insulators (TIs) are said to be stable against non-magnetic impurity scattering due to suppressed backscattering in the Dirac surface states. We solve a lattice model of a three-dimensional TI in the presence of strong potential impurities and find that both the Dirac point and low-energy states are significantly modified: Low-energy impurity resonances are formed that produce a peak in the density of states near the Dirac point, which is destroyed and split into two nodes that move off-center. The impurity-induced states penetrate up to ten layers into the bulk of the TI. These findings demonstrate the importance of bulk states for the stability of TIs and how they can destroy the topological protection of the surface.
\end{abstract}
\maketitle

%
%
Topological insulators (TIs) belong to a new state of matter, which is insulating in the bulk but with a conducting surface, where Dirac-like energy spectra lock spin and momentum together into a spin-helical state.\cite{Hasan10, Qi10}
In strong TIs there are only a single (or odd number of) Dirac surface state, and this band topology protects the surface state against any perturbation that preserves time-reversal symmetry.\cite{Fu07} One simple way to test the topological stability is to probe the spectra in the presence of impurity scattering.
Theoretical results for non-magnetic impurities using a continuum model for the Dirac surface state have indeed shown both the absence of backscattering\cite{Lee09andothers}, as also confirmed experimentally,\cite{Roushan09andothers} and how these impurities never destroy, even locally, the low-energy spectrum, including the Dirac point.\cite{Biswas10} This is in contrast to magnetic impurities, which have been theoretically shown to open a gap in the surface spectrum.\cite{Liu09} Taken together, these results make TIs intriguing candidates for both spintronic devices and topological quantum computation.\cite{Fu08}

We start here with the observation that the argument for topological protection in a TI is based on the first-order scattering amplitude having a node for 180$^\circ$ backscattering. As such, this argument is only a first-order effect, effectively set by the size of the bulk gap. Therefore, we ask a simple but very reasonable question: How stable is the TI surface state against strong perturbations? Since a single impurity, or vacancy, can provide an energy perturbation ($\gtrsim 1$~eV) easily exceeding that of the bulk gap in a TI ($\sim 0.3$~eV), there is no symmetry argument that prevents backscattering from occurring through virtual spin-flip excitations in the bulk. 
Indeed, some recent experimental results have pointed to the importance of bulk-assisted processes, both in terms of linewidth broadening\cite{Park10} and producing localized bound states at defects, not agreeing with results from a surface continuum model.\cite{Alpichshev11, Alpichshev11b} In the latter case, a theoretical analysis of a step defect established that finite surface gradients induce bulk interference.\cite{Alpichshev11}

In this Rapid Communication we investigate the consequences of interplay and coupling between surface and bulk in a TI in the presence of non-magnetic impurities. We do this by studying strong potential impurities in a model three-dimensional (3D) strong TI, explicitly focusing on intragap properties. Some earlier theoretical results exist on a lattice model but they do not discuss any intragap consequences of impurity scattering.\cite{Wang09}
We find that
(i) impurities create localized resonance states which appear at ever lower energies when the impurity strength $U$ is increased, with $E_{\rm res} \simeq  -1/U$. For weak impurities, the resonance peak and the Dirac point are well separated in energy, but the impurities nonetheless move the Dirac point due to an overall effective doping of the system. For strong scatterers the resonance peak moves past the location of the unperturbed Dirac point and, instead, two new Dirac points emerge on both sides of the resonance peak. Thus, the topologically protected Dirac surface-state spectrum with a single apex, or Dirac point, is, at least locally, destroyed by these strong potential scatterers.
(ii) Both the surface state and resonance peak penetrate many ($\gtrsim \! \! 10$) layers into the sample. In combination with a finite bulk gap, this impurity-induced cross-talk between surface and bulk causes the observed disruption of the Dirac spectrum by permitting second-order bulk-assisted scattering processes.
In fact, the resonance states in a TI are similar to states found in both graphene\cite{Pereira06,Wehling07,Ugeda10} and $d$-wave high-temperature superconductors,\cite{Balatsky06} two other materials with Dirac-like low-energy spectra, thus making a strong argument for a unified local response to impurities for all ``Dirac" materials,\cite{Wehling11} once any topological protection is lost.

%
\section{Model}
For a simple, but realistic, model of a strong TI we use a tight-binding model on the diamond lattice with spin-orbit coupling (SOC):\cite{Fu07}
%
\begin{align}
\label{eq:H0}
H_0 = t  \! \sum_{\langle i,j\rangle} \! c^\dagger_{i}c_{j} + \mu \! \sum_i \! c^\dagger_i c_i +
4i\lambda/a^2 \! \! \sum_{\langle \langle i,j\rangle \rangle} \! \! c^\dagger_{i} {\bf s \cdot (d}^1_{ij}\times {\bf d}^2_{ij}) c_{j}.
\end{align}
Here $c_{i}$ is the annihilation operator on site $i$ where we have suppressed the spin index, $t$ the nearest-neighbor hopping, $\mu = 0$ the chemical potential, $\lambda = 0.3t$ the next-nearest-neighbor SOC, $\sqrt{2}a$ the cubic cell size, ${\bf s}$ the Pauli spin matrices, and ${\bf d}_{ij}^{1,2}$ the two bond vectors connecting next-nearest-neighbor sites $i$ and $j$.
In order to access a surface we create a slab of Eq.~(\ref{eq:H0}) along the (111) direction with ABBCC ... AABBC stacking terminations on each side, respectively. We find that for $\gtrsim 5$ lateral unit cells there is minimal cross-talk between the two surfaces.
By further distorting the hopping amplitude to $1.25t$ along one of the nearest-neighbor directions not parallel to (111), this system becomes a strong TI, with a single surface Dirac cone located at one of the $M$ points in the surface Brillouin zone.\cite{Fu07}
We use an energy scale such that the slope of the surface Dirac cone $\hbar v_F\approxeq 1$, which is achieved by setting $t = 2$ throughout this work.
In order to study the effect of a local potential impurity we create a rectangular-shaped surface supercell with $n$ sites along each direction. This gives a supercell surface area of $\sqrt{3}n^2a^2/2$, where we use $a=1$ as the unit of length. We create an impurity on site $i$ on the surface by adding the term $H_{\rm imp} = U c_i^\dagger c_i$, where $U\geq 0$ is the impurity strength, to the Hamiltonian in Eq.~(\ref{eq:H0}). We note that by adding $H_{\rm imp}$ we break particle-hole symmetry and thus our model, even with $\mu =0$, corresponds to a quite general situation.
We solve $H= H_0 + H_{\rm imp}$ in the supercell by using exact diagonalization. We find that a $50 \times 50$ $k$-point grid gives sufficient resolution, while at the same time using a Gaussian broadening of $\sigma = 0.005$ when calculating the local density of states (LDOS).

%
\section{Dependence on U}
\begin{figure}[htb]
\includegraphics[scale = 0.8]{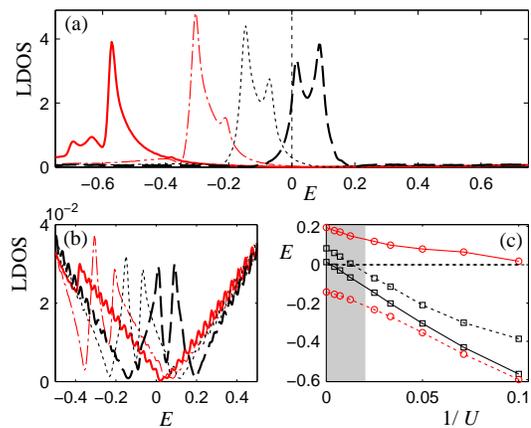}
\caption{\label{fig:Usweep} (Color online) (a) Average LDOS per energy and area unit on nearest-neighbor surface sites for $U = $10 (thick red), 20 (dashed-dotted red), 40 (dotted black) impurity and vacancy, $U\rightarrow \infty$ (thick dashed black) for $n = 10$. (b) LDOS far from impurity. (c) Dependence on $1/U$ for peak position (black $\square$) and Dirac point, (red $\circ$). Dashed lines mark the position of the right-hand side subpeak and left-hand side Dirac point, with the gray shaded area indicating where the latter is present as a zero DOS point. Horizontal and vertical thin dashed lines indicate $E = 0$.
}
\end{figure}

We first focus on the change in the LDOS as a function of impurity strength $U$. Figure \ref{fig:Usweep}(a) shows the LDOS on the nearest-neighboring surface sites to the impurity for a sequence of different values of $U$ at fixed impurity concentration. For very weak impurities ($U\lesssim10$) the impurity resonance resides inside the bulk valence-band. With increasing $U$, the resonance peak moves to lower energies and enters the bulk gap region, where it  becomes clearly visible amid the Dirac surface spectrum. Eventually, for large enough $U$, the resonance peak moves {\it past} $E = 0$, where the Dirac point of the unperturbed system is located. With increasing $U$ the peak does not significantly change its height or width, although a second subpeak develops on the right-hand side for large $U$ values. This subpeak even comes to slightly dominate the original peak when $U \rightarrow \infty$. We believe the origin of the double peak is the overlap of impurity states at large impurity concentration, when the impurity states start forming a band and the resonance peak broadens due to overlap. Indeed, the peak-peak splitting goes down when increasing $n$ [see Fig.~\ref{fig:Uvac}(a)]. The whole peak structure is non-dispersive in energy over the whole supercell and in Fig.~\ref{fig:Usweep}(c) we see that the position of the impurity resonance scales as $E_{\rm res} \simeq -1/U$.
In a two-dimensional (2D) continuum model for the surface state, the same $U$ dependence was established but the resonance peak was found to get narrower and taller with increasing impurity strength and it finally disappears at unitary scattering.\cite{Biswas10}
We further find that the peak height decays with distance approximately as $1/R^{-3}$ for all values of $U$, a faster decay than found previously.\cite{Biswas10} This difference could be related either to the inherent 3D nature of the state or to the fact that we are only looking at the short distance behavior.
%
The finite resonance peak at low energies for strong scattering raises the question of stability of the Dirac point.  Figure~\ref{fig:Usweep}(b) shows the LDOS at surface sites far from the impurity, and we see that with increasing scattering strength, the apex of the Dirac spectrum, i.e.,~the Dirac point, moves to positive energies and that the resonance peak never moves past this point. However, for large $U$ we also see the development of a {\it second} Dirac point on the left-hand side of the impurity peak, such that at unitary scattering the negative-energy Dirac cone terminates in a point to the left of the resonance peak, whereas the positive-energy Dirac cone terminates in a point to the right of the peak. Thus, for very large $U$, the resonance peak is situated between the apices of the valence- and conduction-band Dirac cones. 
In Fig.~\ref{fig:Usweep}(c) the development of the two Dirac points with respect to $1/U$ is plotted in red (light gray). The shaded gray area indicates approximately the region of large $U$ where the second, left-hand side, Dirac point is present. Beyond this region there is no real second Dirac point, only a non-zero dip in the DOS. 
Below we will in detail analyze separately the case of weak scattering, when the resonance peak and the Dirac point are well separated, and the case of strong scattering, when the peak structure is in close proximity to the Dirac point.

%
%
\section{Weak impurities}
For a weak impurity we find that the resonance peak moves to slightly lower energies when the impurity concentration decreases. At the same time the right-hand side subpeak clearly diminishes, such that the center of mass of the peak is still approximately constant. 
More distinct is the development of the position of the Dirac point when changing impurity strength and concentration.
Not only does he Dirac point shift to positive energies, but also the high-energy feature of the DOS closes linearly to the, now shifted, Dirac point.
With the overall chemical potential $\mu$ set to zero, one could imagine an overall shift in the Dirac point directly related to $U/n^2$, as there are $n^2$ surface sites and $U$ acts as a local doping source. Now, this argument is too simplistic since it does not, e.g.,~take into account any of the sub-surface sites, or how the effective doping $U$ spreads over the surface. Despite this, we still see a shift of the Dirac point to positive energies, which is approximately linear in $U$. Moreover, for $U = 14$ the Dirac point position is roughly proportional to $n^{-1.5}$, whereas for $U = 40$ the concentration dependence has weakened to $n^{-1.1}$. We therefore attribute the shift of the Dirac point to positive energies to an effective and uniform surface doping by the impurities, which, in the limit of an isolated impurity, disappears.

%
\section{Strong impurities}
\begin{figure}[htb]
\includegraphics[scale = 0.8]{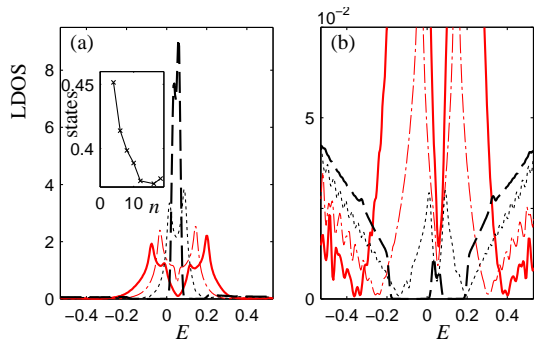}
\caption{\label{fig:Uvac} (Color online) (a) Average LDOS per energy and area unit on nearest-neighbor surface sites to a vacancy for supercell sizes $n = 4$ (thick red), 6 (dashed-dotted red), 10 (dotted black), and 18 (thick dashed black). The inset shows the resonance peak weights (states per area unit) as function of supercell size $n$. (b) LDOS as far away from the impurity as possible. The finite gap for $n = 18$ is due to only three lateral unit cells.
}
\end{figure}

Now, let us concentrate on the strong scattering regime, where the resonance peak and the Dirac point can no longer be thought of as separate entities. In Fig.~\ref{fig:Uvac} we plot the LDOS for a vacancy ($U\rightarrow \infty$)  at different impurity concentrations. We see directly in Fig.~\ref{fig:Uvac}(a) that for decreasing impurity concentration, the impurity resonance becomes significantly sharper, the impurity peak-peak splitting diminishes, and the prominent dip in between the two peaks, at around $E = 0.5$, disappears.
The inset further shows how the total number of states within the peak does not significantly change with $n$ and also seems to converge to a value of $\sim 0.4$ states per unit area for low concentrations.
Thus, for the dilute impurity case $n \rightarrow \infty$, we expect a single resonance peak to be present in the low-energy spectrum, contrary to 2D surface-state continuum model results.\cite{Biswas10} We attribute this difference to the 3D dimensionality of the problem.
%
%
In Fig.~\ref{fig:Uvac}(b) we plot the LDOS as far away from the impurity as the supercell construction allows, in order to focus on the development of the Dirac points with impurity concentration. These Dirac points are located on both sides of the resonance peak, symmetrically with respect to the center of the peak structure. These are not only the points where the DOS reaches down to zero, but they are also the points where the high-energy negative and positive Dirac cone spectra close, i.e.,~where the Dirac apices associated with the slopes at higher energies are located. This verifies the nature of these two points as Dirac points. With decreasing impurity concentrations the two Dirac points move closer together since the resonance peak then becomes narrower. Thus, for very low impurity concentrations the apices of the negative and positive high-energy spectra will almost converge to a single point at the center of the resonance peak.
The finite gap in the lowest concentration sample ($n = 18$) is due to cross-talk between the two surfaces of the slab, because of computational limitations in the total system size. The gap region covers what would have been a finite DOS due to both the resonance peak and the Dirac surface spectrum. We have carefully checked that such a finite cross-talk gap does not change either the resonance peak weight or the high-energy spectrum features.
We also note that with decreasing impurity concentration the high-energy spectrum becomes smoother. In the lowest concentration sample, the high-energy Dirac spectrum has become smooth enough to distinguish a clear kink at around $E = \pm 0.3$. Below this kink the linear spectra on each side close at the two Dirac points at the base of the resonance peak. However, above the kink, both the positive and negative linear Dirac spectra instead have their apices at $E = 0$. Moving closer to the impurity site, the kink position does not change, but the slope above, at higher energies, starts to coincide with the slope below, at lower energies.
Therefore, close to the impurity, the two Dirac points at the base of the resonance peak are still the only defining parameters for the whole energy spectrum of the surface band. This is true for any impurity concentration, including the case of an isolated impurity.
However, far from the impurity, the surface-state spectrum located above a kink is ``healed" back to its unperturbed state, i.e. its slope indicates closing at a single Dirac point at $E = 0$. We find that the kink moves to somewhat lower energies as the impurity concentration decreases, such that  this ``healing" above the kink takes place at lower energies for more dilute impurities. Since we are unable, at the present moment, to model significantly larger systems than those reported in Fig.~\ref{fig:Uvac}, we cannot say anything definitive about the low-energy spectrum far away from a single isolated impurity. This will depend on the kink position as the impurity concentration decreases. It would be reasonable to expect that the kink eventually reaches down toward $E = 0$ for an isolated impurity, thus leaving little, if any, trace of the impurity resonance and its double Dirac points very far from an isolated impurity in any part of the energy spectrum.
%
%
\section{Bulk penetration}
\begin{figure}[htb]
\includegraphics[scale = 0.8]{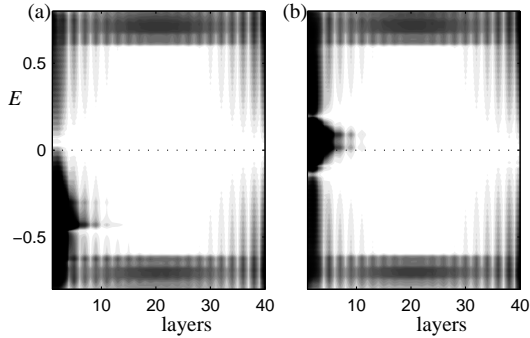}
\caption{\label{fig:bulk} Average LDOS on nearest-neighbor sites to a $U = 7$ impurity (a) and a vacancy (b) for $n = 10$ plotted for each layer across a seven lateral unit cell wide slab. Zero (white), 0.1 (black) states per energy and area unit. The dotted line marks $E = 0$.
}
\end{figure}

Above we have established that a large resonance peak resides inbetween two emerging Dirac points for strong potential  impurities. Thus, the topologically protected Dirac surface-state spectrum with a single Dirac point is, at least locally, destroyed by such strong potential scatterers. This might seem surprising as the surface-state is topologically protected from any perturbation that is time-reversal invariant. However, often forgotten in this line of reasoning is both the finite size of the bulk gap and the fact that the surface and bulk states have a finite spatial overlap in any realistic TI.
Together these two effects open up the possibility of virtual excitations to the bulk of the TI. With spin flips allowed in the bulk, virtual spin-flip bulk excitations give rise to (the otherwise absent) backscattering on the surface and thus the topological protection of the Dirac point is lost.
Since the electrostatic energy for strong scatterers can easily surpass that of the bulk gap in real TIs, these bulk-assisted processes can become very common, granted that there is also a sizable spatial overlap between the bulk and surface states.
In Fig.~\ref{fig:bulk} we plot the LDOS on nearest-neighbor sites to the impurity across all layers of the slab for both a weak impurity [Fig.~\ref{fig:bulk}(a)] and a vacancy [Fig.~\ref{fig:bulk}(b)].
%
On the opposite surface (layer 40) we see the LDOS pattern of an unperturbed Dirac cone centered at the chemical potential $\mu = 0$. This state slowly joins the bulk states, present at $E\approx \pm 0.6$, when penetrating into the slab. At the 10-layer depth the remnant DOS from the surface state is less than 0.002 and is also only located at energies close to the bulk gap.
On the surface with the impurity (layer 1) we see how the resonance peak significantly modifies the spectrum. A weak impurity markedly enhances the LDOS around its resonance peak, here at $E\approx -0.4$, but, to some degree, the LDOS is enhanced over the whole energy spectrum in the surface layer. At energies far from the resonance peak, the unperturbed Dirac spectrum is recovered in subsurface layers. However, close to the resonance peak the LDOS is enhanced even deep below the impurity, with traces of the peak found somewhat deeper than the unperturbed surface state.
For a vacancy the bulk penetration is rather similar, but with the notable difference that the resonance peak now sits close to $E = 0$ and two new Dirac points have emerged on either side of this peak.
Thus, our results for a general model of a 3D TI show that both a finite bulk gap and a finite and sizable penetration depth for the surface state (and the resonance peak) are present, together giving rise to bulk-assisted scattering.
Since the argument on topological protection for the surface state relies on suppressed 180$^\circ$ backscattering in the case of 2D scattering, it totally ignores the contributions of 3D scattering processes that engage bulk states.
Therefore, there is no inherent protection left for the original Dirac point from strong non-magnetic impurities, and the resonance peak at low energies, with the accompanied two Dirac points, are simply a consequence of a bulk-surface interaction in the system.
These results show that any realistic solution of the impurity problem in TIs has to include a realistic calculation of the bulk contribution, and consequently, the argument for suppressed backscattering also needs to be modified to include bulk-assisted scattering.
We also note that large impurity resonance peaks close to the Fermi level are going to be sensitive to strong Coulomb interactions, which can lead to spin-polarized splitting and thus spontaneous local time-reversal symmetry breaking. However, we leave a detailed analysis of interaction effects for future work.

\section{Note added} At this point we would also like to note that scanning tunneling spectroscopy (STS) results on non-magnetic impurities in Bi$_2$Se$_3$ have recently appeared, confirming the existence of strong resonance states at the Dirac point.\cite{Teague12} 
%
\begin{acknowledgments}
We are grateful to Z.~Alpichshev,  R.~Biswas, T.~Hanaguri,  D.~H. Lee, A.~Kapitulnik, H.~Manoharan, T.~Wehling, and S.~C.~Zhang  for discussions. AMBS acknowledges support from the Swedish research council (VR). Work at Los Alamos was supported by U.S.~DOE Basic Energy Sciences and in part by the Center for
Integrated Nanotechnologies, operated by LANS, LLC,
for the National Nuclear Security Administration of the U.S.
Department of Energy under Contract No.~DE-AC52-06NA25396.
\end{acknowledgments}


\end{document}